TITLE:

# Photonic Quantum Computation with Waveguide-Linked Optical Cavities and Quantum Dots

Makoto Yamaguchi[*], Takashi Asano[*], Yoshiya Sato and Susumu Noda[†]

*Department of Electronic Science and Engineering, Kyoto University,*

*Katsura, Nishikyo-ku, Kyoto 615-8510, Japan*

PACS numbers: 42.50.Pq, 42.50.Ct, 73.21.La, 78.67.Hc

**ABSTRACT:**

We propose a new scheme for solid-state photonic quantum computation in which trapped photons in optical cavities are taken as a quantum bit. Quantum gates can be realized by coupling the cavities with quantum dots through waveguides. The proposed scheme allows programmable and deterministic gate operations and the system can be scaled up to many quantum bits.

Quantum information processing (QIP) promises significant improvements in computational performance and communication security [1]. Many physical systems have been studied for the development of quantum bits (qubits) and necessary gate operations, including ion traps [2], quantum dots [3], Josephson junctions [4], and nuclear magnetic resonances [5]. Qubits encoded in photonic quantum states (photonic qubits) are one possible candidate for QIP, and are currently the only realistic carriers for long-distance quantum communication because of their low decoherence and transmission speed of light. In this context, photonic qubits are also attractive for quantum computation (QC). However, the application of photonic qubits faces two significant challenges. First, optical nonlinearities are generally too small at the single-photon level to achieve the interaction between two photons that is needed, for example, in a controlled-NOT (CNOT) gate. Second, the photonic qubits are essentially flying, making operations under programmable conditions difficult. Although intensive efforts are currently being devoted to overcome these difficulties, including implementation of the probabilistic KLM (Knill-Laflamme-Milburn) scheme [6], the use of an optical cavity to enhance atom-photon interactions [7], and the use of one-way QC involving cluster states [8], these challenges remain daunting with respect to the practical application of photonic qubits despite proof-of-principle demonstrations.

Here, we propose a novel cavity-based photonic qubit where programmable and deterministic control is possible. Our photonic qubit is encoded in two distant optical cavities linked by a waveguide (WG), as shown in Fig. 1 [9]; this system is motivated by recent progress in the development of on-chip photonic devices, such as photonic crystal (PC) cavities [10], micro-toroidal cavities [11], dynamic control of the quality ($Q$) factor [12], and cavity quantum electrodynamics using a single quantum dot (QD) [13]. In our scheme, the gate operations are achieved by dynamic control of the optical length (or refractive index) of the WG. The interaction between the two photons is mediated by a single QD (which can be treated in general as a V-type three-level system) that is embedded in the intersection of two orthogonal WGs (Fig. 2). Our scheme simultaneously permits a number of significant advances to be made. (i) It enables programmable operations because the optical length of the WGs can be controlled dynamically after determination of the device structure [12]. (ii) The single QD allows a deterministic CNOT operation to be achieved. (iii) Our system can be integrated and scaled up to many qubits with properly designed WGs [14], providing a scalable platform for QC by itself. (iv) Finally, each unit described below is compatible with current photonic QIP schemes and could also play a key role in improving their performance. Even though practical application of our scheme may require further developments of current technologies and designs, our proposal will call for and accelerate them for future implementation of on-chip photonic QIP.

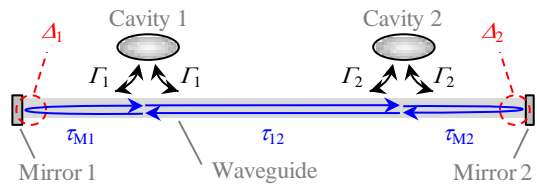

FIG. 1 (color online). Schematic picture of the proposed system for one-qubit operations.

First, we describe the ways in which encoding and one-qubit gate operations are performed by discussing the effective Hamiltonian of the system shown in Fig. 1, where the WG is evanescently coupled with the cavities. The initial Hamiltonian for this system can be expressed as

$$\hat{H} = \hbar \sum_{l=1}^{2} \omega_{cl} \hat{a}_{cl}^{\dagger} \hat{a}_{cl} + \hbar \sum_{\lambda \in \{FP\}} \omega_{\lambda} \hat{a}_{\lambda}^{\dagger} \hat{a}_{\lambda} + i\hbar \sum_{l=1}^{2} \sum_{\lambda \in \{FP\}} [g_{l,\lambda} \hat{a}_{cl} \hat{a}_{\lambda}^{\dagger} - g_{l,\lambda}^{*} \hat{a}_{\lambda} \hat{a}_{cl}^{\dagger}], \tag{1}$$

where $g_{l,\lambda} \equiv g_{l,\lambda}^{L} + g_{l,\lambda}^{R}$ is the coupling constant between the cavity modes and the Fabry-Perot (FP) modes formed in the WG. Here, $\hat{a}_{cl}$ denotes the $l$-th cavity-mode annihilation operator and $\hat{a}_{\lambda}$ is the $\lambda$-th FP mode annihilation operator. As shown in the supplementary material [15], this Hamiltonian can be derived as the quantization of classical coupled mode theory involving all of the input-output relations between incoming and outgoing waves. The absolute values of $g_{l,\lambda}^{i}$ ($i$ = L, R and $l$ = 1, 2) are expressed as $|g_{l,\lambda}^{i}| = \sqrt{\Gamma_{l}/\tau_{P}}$, where $2\Gamma_{l}$ is the sum of the photon emission rates from the $l$-th cavity into the right-hand and left-hand directions of the WG and $\tau_{P}$ is the time taken for the light to propagate along the WG and back ($\tau_{P} = \tau_{M1} + \tau_{M2} + 2\tau_{12}$), as shown in Fig. 1. The phase differences of $g_{l,\lambda}^{i}$ are given by $\arg[g_{2,\lambda}^{L}] - \arg[g_{2,\lambda}^{R}] = \omega_{\lambda} \tau_{M2} + \Delta_{2}$ and $\arg[g_{1,\lambda}^{L}] - \arg[g_{2,\lambda}^{L}] = \arg[g_{2,\lambda}^{R}] - \arg[g_{1,\lambda}^{R}] = \omega_{\lambda} \tau_{12}$, where $\Delta_{l}$ is the change in phase that occurs when a photon is reflected at the $l$-th mirror and $\tau_{M1}$, $\tau_{M2}$, and $\tau_{12}$ are the propagation times of light between the cavities. When both cavities have the same properties (i.e., resonant frequency $\omega_{c1} = \omega_{c2} \equiv \omega_{0}$ and emission rate $\Gamma_{1} = \Gamma_{2} \equiv \Gamma_{c}$), elimination of the waveguide's degrees of freedom [15] gives the following effective Hamiltonian:

$$\hat{H}_{I}^{eff} = \hbar \sum_{l=1}^{2} \omega_{cl}^{eff} \hat{a}_{cl}^{\dagger} \hat{a}_{cl} + \hbar g_{12}^{eff} [\hat{a}_{c1}^{\dagger} \hat{a}_{c2} + \text{h.c.}], \tag{2}$$

where the effective resonant frequencies are

$$\omega_{c1}^{eff} \equiv \omega_0 + \chi(\theta_{M1}, 2\theta_{12} + \theta_{M2}, \theta_P)\Gamma_c, \qquad (3)$$

$$\omega_{c2}^{eff} \equiv \omega_0 + \chi(\theta_{M2}, 2\theta_{12} + \theta_{M1}, \theta_P)\Gamma_c, \qquad (4)$$

and the effective coupling constant is

$$g_{12}^{eff} \equiv \chi(\theta_{M1}, \theta_{M2}, \theta_P)\Gamma_c. \qquad (5)$$

Here, $\theta_{M1} \equiv \omega_0 \tau_{M1} + \Delta_1$, $\theta_{M2} \equiv \omega_0 \tau_{M2} + \Delta_2$, and $\theta_{12} \equiv \omega_0 \tau_{12}$ denote the phase differences caused by the propagation of light; $\theta_P$ and the function $\chi$ are defined as $\theta_P \equiv \theta_{M1} + \theta_{M2} + 2\theta_{12}$ and $\chi(x, y, z) \equiv [\cos([x+y]/2)+\cos([x–y]/2)]/\sin(z/2)$, respectively. In the derivation of Eqs. (2)-(5), it is assumed that the propagation time for the round trip of light along the WG is much shorter than the inverse photon emission rate ($\tau_P \ll 1/2\Gamma_c$), and that $\omega_0$ is sufficiently detuned from the FP modes ($|\omega_0 - \omega_\lambda| \gg |g_{l,\lambda}|$ for $l = 1, 2$). In this limit, the effective Hamiltonian in Eq. (2) can be interpreted as an interaction between the cavities mediated by a virtual photon of the FP modes. It is apparent from Eqs. (3)-(5) that $\omega_{c1}^{eff}$, $\omega_{c2}^{eff}$ and $g_{12}^{eff}$ can be controlled by the phase differences $\theta_{M1}$, $\theta_{M2}$, and $\theta_{12}$, which can be varied by tuning $\tau_{M1}$, $\tau_{M2}$, and $\tau_{12}$; this can be achieved by dynamic control of the optical length (or the refractive index) of the WG [12]. Therefore, programmable one-qubit gate operations can be performed when a logical state $|a\rangle_L$ ($a = 0, 1$) is encoded by a physical state $|a\rangle_{c1}|1-a\rangle_{c2}$, where $|a\rangle_{cl}$ denotes $a$ photons in the $l$-th cavity mode. Here, the required operations for the one-qubit gate can be divided into three types of control: holding the qubit, inducing Rabi oscillations, and controlling the relative phase between $|1\rangle_L$ and $|0\rangle_L$. First, one can hold the states of the qubit by setting $\omega_{c1}^{eff} = \omega_{c2}^{eff}$ and $g_{12}^{eff} = 0$ in Eq. (2), a condition that can be achieved when the phase differences are set to $\theta_{M1} = \pi$, $\theta_{M2} = 0$, and $\theta_{12} = 0$. Second, Rabi oscillations can be induced under the conditions of $\omega_{c1}^{eff} = \omega_{c2}^{eff}$ and $g_{12}^{eff} = \Gamma_c$, which can be realized by setting $\theta_{M1} = \pi/2$, $\theta_{M2} = \pi/2$, and $\theta_{12} = 0$. Finally, control of the relative phase between $|1\rangle_L$ and $|0\rangle_L$ is achieved when $\omega_{c2}^{eff} - \omega_{c1}^{eff} = \Gamma_c$ and $g_{12}^{eff} = 0$, conditions that are obtained by setting $\theta_{M1} = \pi$, $\theta_{M2} = \pi/2$, and $\theta_{12} = -\pi/4$. These three types of control can be used to construct an arbitrary state of the single qubit, allowing a complete set of one-qubit gate operations.

The arguments above show that programmable one-qubit gate operations are possible for our photonic qubit. However, a major difficulty facing the use of photonic qubits is the achievement of non-trivial two-qubit gate operations. Therefore, we next explain how the CNOT operation can be performed using a single QD (a V-type three-level system), which transforms a two-qubit logical state $|a\rangle_{L1}|b\rangle_{L2}$ ($a$, $b$ = 0, 1) into $|a \oplus b\rangle_{L1}|b\rangle_{L2}$. This operation can be constructed from a quasi-CNOT (q-CNOT) operation $|a\rangle_{L1}|b\rangle_{L2} \rightarrow (-1)^b |a \oplus b\rangle_{L1}|b\rangle_{L2}$ by utilizing a one-qubit operation $|b\rangle_{L2} \rightarrow (-1)^b |b\rangle_{L2}$. Therefore, in this paper we simplify the system by adopting the q-CNOT gate as a standard two-qubit gate instead of the CNOT gate. The system required for our q-CNOT operation $|a\rangle_{L1}|b\rangle_{L2}|g\rangle \rightarrow (-1)^b |a \oplus b\rangle_{L1}|b\rangle_{L2}|g\rangle$ is shown in Fig. 2, where the logical state $|a\rangle_{L1}|b\rangle_{L2}$ is encoded by $|a\rangle_{c3}|1-a\rangle_{c4}|b\rangle_{c5}|1-b\rangle_{c6}$ (the first and second qubits correspond to the target and control qubits, respectively). The following three assumptions are required. First, the energy levels of the $|x\rangle$ and $|y\rangle$ states in the QD are degenerate. Second, the $|y\rangle$-$|g\rangle$ optical transition is allowed only for WG1 and the $|x\rangle$-$|g\rangle$ transition is allowed only for WG2 because these transitions are polarized orthogonally in the x and y directions, respectively. Finally, the photon emission rates of the third and fourth (fifth and sixth) cavities into the WGs are designed to be $2\Gamma$ ($4\Gamma$) when that of the QD is $4\Gamma$. We will now discuss the way in which the q-CNOT operation for the encoded state $|a\rangle_{c3}|1-a\rangle_{c4}|b\rangle_{c5}|1-b\rangle_{c6}$ is performed.

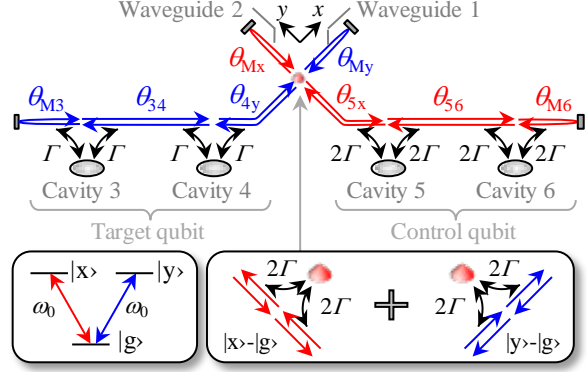

FIG. 2 (color online). Schematic picture of the proposed system for two-qubit operation. The QD is treated as a V-type three level system (left inset) where the |x⟩-|g⟩ transition is polarized in the x-direction and the |y⟩-|g⟩ transition is polarized in the y-direction. Therefore, each QD transition is allowed only for one of the orthogonal WGs (right inset). The eight $\theta$'s are the phase differences caused by the propagation of light.

The q-CNOT operation consists of the following three steps. (A) The information of the control qubit is transferred once and stored in the QD. (B) The state of the QD determines whether the logical state in the target qubit is flipped or held. (C) The information stored in the QD is returned to the control qubit. These steps can be performed by controlling the eight phase differences shown in Fig. 2 in a similar manner to the phase differences in the one-qubit operations described above (the specific values of the phase differences in each step are summarized in the supplementary material [15]). Here, we give detailed explanations of the principles of operation of each step. In the first step (A), only the fifth cavity and the QD are coupled through WG2; the other elements are decoupled [15]. Under this condition, the evolution of the system can be described by a unitary operator $\hat{U}_1^t \equiv \exp(-i\hbar^{-1}\hat{H}_1 t)$ with $\hat{H}_1 \equiv \hat{H}_0 + \hat{V}_1$, where $\hat{H}_0 \equiv \hbar\omega_0 \sum_{l=3}^{6} \hat{a}_{cl}^\dagger \hat{a}_{cl} + \hbar\omega_0[\hat{\sigma}_{x,x} + \hat{\sigma}_{y,y}]$ denotes the original eigen-energy of each state. In addition,

$$\hat{V}_1 \equiv 2\hbar\Gamma[\hat{a}_{c5}^\dagger \hat{a}_{c5} + \hat{\sigma}_{x,x}] - 2\hbar\Gamma[\hat{a}_{c5}^\dagger \hat{\sigma}_{g,x} + \text{h.c.}] \tag{6}$$

describes the detuning from $\omega_0$ and the interactions induced by WG2 ($\hat{\sigma}_{i,j}$ is defined as $|i\rangle\langle j|$ for $i, j =$ g, x, y). Therefore, the unitary operator $\hat{U}_1^{\pi/4\Gamma}$ can convert a photon stored in the fifth cavity into the |x› state by means of the Rabi oscillations ($|a\rangle_{L1}|1\rangle_{c5}|0\rangle_{c6}|g\rangle \rightarrow |a\rangle_{L1}|0\rangle_{c5}|0\rangle_{c6}|x\rangle$). In contrast, no such conversion occurs when there is no photon in the fifth cavity ($|a\rangle_{L1}|0\rangle_{c5}|1\rangle_{c6}|g\rangle \rightarrow |a\rangle_{L1}|0\rangle_{c5}|1\rangle_{c6}|g\rangle$). As a result, the information of the control qubit can be stored in the QD. In the second step (B), the evolution of the system can be described by a unitary operator $\hat{U}_2^t \equiv \exp(-i\hbar^{-1}\hat{H}_2 t)$, with $\hat{H}_2 \equiv \hat{H}_0 + \hat{V}_2$ and

$$\hat{V}_2 \equiv \hbar\Gamma \sum_{l=3}^{4} \hat{a}_{cl}^\dagger \hat{a}_{cl} + 2\hbar\Gamma\hat{\sigma}_{y,y} + \hbar\Gamma[\hat{a}_{c3}^\dagger \hat{a}_{c4} + \text{h.c.}] - \sqrt{2}\hbar\Gamma\sum_{l=3}^{4}[\hat{a}_{cl}^\dagger \hat{\sigma}_{g,y} + \text{h.c.}], \tag{7}$$

which can be achieved by coupling the third and fourth cavities and the QD simultaneously through WG1 [15]. In the case where the QD is in the |g› state, the third and fourth cavities and the QD can interact with each other simultaneously. This type of evolution can be derived from Eq. (7) and is expressed as

$$\hat{U}_2^t |a\rangle_{L1}|0\rangle_{c5}|1\rangle_{c6}|g\rangle = \frac{\exp(-i2\omega_0 t)}{\sqrt{2}} \left( \frac{1+\exp(-i4\Gamma t)}{2}|+\rangle_{L1}|g\rangle + [2a-1]|-\rangle_{L1}|g\rangle \right. \\ \left. + \frac{1-\exp(-i4\Gamma t)}{2}|0\rangle_{c3}|0\rangle_{c4}|y\rangle \right) |0\rangle_{c5}|1\rangle_{c6}, \tag{8}$$

where $|\pm\rangle_{L1} \equiv [|1\rangle_{L1} \pm |0\rangle_{L1}]/\sqrt{2}$. Therefore, the initial state $|a\rangle_{L1}|0\rangle_{c5}|1\rangle_{c6}|g\rangle$ can be recovered after a time $T = \pi/2\Gamma$, except for a factor $\exp(-i2\omega_0 t)$, which is a meaningless global phase. In contrast, the QD does not interact with the third and fourth cavities when it is in the |x› state because the |x›-|g› transition is not allowed for WG1. In this case, simple Rabi oscillations with a period $T_{\text{Rabi}}^{34} = \pi/\Gamma$ are induced between the cavities, the evolution of which can be expressed by

$$\hat{U}_2^t |a\rangle_{L1} |0\rangle_{c5} |0\rangle_{c6} |x\rangle = \frac{\exp(-i2\omega_0 t)}{\sqrt{2}} \left( \exp(-i2\Gamma t) |+\rangle_{L1} + [2a-1]|-\rangle_{L1} \right) |0\rangle_{c5} |0\rangle_{c6} |x\rangle. \quad (9)$$

Therefore, the logical state of the target qubit is flipped after a time $T_{\text{Rabi}}^{34}/2$. It is thus possible to flip the target qubit when the QD is set in the |x› state and to hold the qubit when it is in the |g› state after a time $\pi/2\Gamma$ because the special condition of $T = T_{\text{Rabi}}^{34}/2$ is now satisfied. We note (see Fig. 2) that this ingenious operation is enabled by photon emission rates of $4\Gamma$ for the QD and $2\Gamma$ for the third and fourth cavities. Thus, the unitary operator in this step is described by $\hat{U}_2^{\pi/2\Gamma}$. In the final step (C), the fifth cavity and the QD are again coupled through WG2 in exactly the same fashion as in step (A) (|a›$_{L1}$|0›$_{c5}$|0›$_{c6}$|x› → |a›$_{L1}$|1›$_{c5}$|0›$_{c6}$|g›, |a›$_{L1}$|0›$_{c5}$|1›$_{c6}$|g› → |a›$_{L1}$|0›$_{c5}$|1›$_{c6}$|g›). Therefore, the evolution of the system is again given by the unitary operator $\hat{U}_1^{\pi/4\Gamma}$. The states that are transformed by the three unitary operations are described in Fig. 3, from which it is apparent that the combined unitary operation $\hat{U}_1^{\pi/4\Gamma} \hat{U}_2^{\pi/2\Gamma} \hat{U}_1^{\pi/4\Gamma}$ is equivalent to a deterministic q-CNOT operation. Thus, these three steps can also provide a deterministic CNOT operation using a one-qubit operation.

```
                    Û₁^(π/4Γ)                          Û₂^(π/2Γ)                          Û₁^(π/4Γ)
|0›c3|1›c4|0›c5|1›c6|g›  →  |0›c3|1›c4|0›c5|1›c6|g›  →  |0›c3|1›c4|0›c5|1›c6|g›  →  |0›c3|1›c4|0›c5|1›c6|g›
|1›c3|0›c4|0›c5|1›c6|g›  →  |1›c3|0›c4|0›c5|1›c6|g›  →  |1›c3|0›c4|0›c5|1›c6|g›  →  |1›c3|0›c4|0›c5|1›c6|g›
|0›c3|1›c4|1›c5|0›c6|g›  →  |0›c3|1›c4|0›c5|0›c6|x›  →  −|1›c3|0›c4|0›c5|0›c6|x›  →  −|1›c3|0›c4|1›c5|0›c6|g›
|1›c3|0›c4|1›c5|0›c6|g›  →  |1›c3|0›c4|0›c5|0›c6|x›  →  −|0›c3|1›c4|0›c5|0›c6|x›  →  −|0›c3|1›c4|1›c5|0›c6|g›
              Step (A)                          Step (B)                          Step (C)
```

FIG. 3. Transformations of physical states by the unitary operations of the q-CNOT gate. The global phase that the states have in common is omitted.

We now focus on the gate-operation time and the effect of decoherence. The one-qubit operation time (~ $\pi/\Gamma_c$) can be engineered from a minimum of a few picoseconds by changing the distance between the WG and cavities [12]. In contrast, the q-CNOT operation time ($\pi/\Gamma$) is limited by the photon emission rate from the QD into the WGs ($4\Gamma$) because the QDs are less controllable than the cavities and the QD emission rate condition above must also be satisfied. The q-CNOT operation time is estimated to be ~ 484 ps ($1/4\Gamma = 38.5$ ps) when the group index $n_g$ is increased up to ~ 200 in the WGs [16]. Thus, the q-CNOT gate is much slower than the one-qubit gates. Decoherence in our system is dominated by pure-dephasing in the QD and photon loss from the cavities into free space. The fidelity $F$ for the q-CNOT gate can be calculated from the quantum master equation [13] by varying the pure-dephasing rate $2\gamma_{phase}$ (Fig. 4(a)) and the $Q$-factors of the cavities (Fig. 4(b)). From Fig. 4(a), it can be seen that $2\hbar\gamma_{phase} < 2$ μeV is required when $F > 0.9$ is desired, which is achievable because $2\hbar\gamma_{phase} \sim 2$ μeV has been reported at cryogenic temperature [17]. Figure 4(b) shows that $Q > 2.0\times10^7$ is necessary for $F > 0.9$, a value that has been theoretically attained in designed PC cavities but which is higher than experimentally realized $Q$-factors (currently $Q < 2.5\times10^6$) [10]. Several solutions can be considered to overcome this difficulty. One approach is to use micro-toroidal cavities with ultra-high $Q$-factors ($> 1.0\times10^8$) [11], and another is to employ slot-type WGs [18] in order to increase the operation speed that is limited by the emission rate $4\Gamma$ of the QD. It would also be beneficial to detect photons that are lost into free space and to eliminate such data as failures when detected during operations.

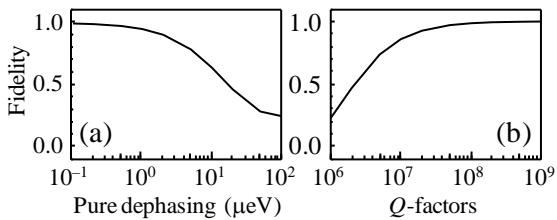

FIG. 4. Fidelity of the q-CNOT gate as a function of (a) pure dephasing and (b) the $Q$-factors of the cavities. The other parameter required in this calculation is $1/4\Gamma = 38.5$ ps.

Finally, we discuss the external ports of our systems, which are reached through additional WGs. The systems presented can both emit and trap photons via dynamic control of the $Q$-factors [12], as shown in Fig. 5. Therefore, individual units can be linked to each other and connected to the external world on demand with the directional switching of flying photons [14], which enables scalability (one possible structure for many qubits is discussed in the supplementary material [15]). The systems can also be initialized because photons can be introduced from the outside. Alternatively, it is possible to utilize the QD in the two-qubit gate (Fig. 5(b)) as a single photon source [15], which can simplify the process of initialization. Furthermore, our design is compatible with current QIP schemes [6, 8] because it is relatively simple for photonic qubits to be transferred from one scheme to another. One such example is apparent in Fig. 5(a) where the propagating photon in the WGs is equivalent to the dual-rail qubit. Therefore, our systems are also applicable as devices to improve the performance of current photonic QIP. In particular, the system shown in Fig. 5(b) can act as a highly efficient entangled photon source by itself, which is essential for the present KLM scheme, one-way QC, and quantum communication.

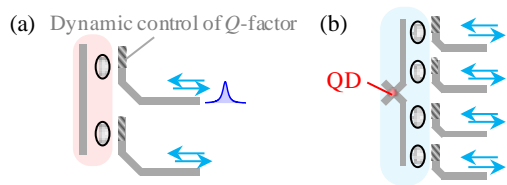

FIG. 5 (color online). External connections for (a) one-qubit and (b) two-qubit gates.

In summary, we have described a new QIP scheme based on waveguide-linked optical cavities and QDs. The proposed scheme enables programmable and deterministic operations and provides a scalable platform for quantum computation, thus forming an important step toward the future implementation of on-chip photonic QIP.

We thank K. Kojima, T. Kojima, T. Nakamura, and K. Hikoyama for helpful discussions. This work is partly supported by Research Programs (Grant-in-Aid, Global Center of Excellence, and Core Research for Evolutional Science and Technology) for Scientific Research from the Ministry of Education, Culture, Sports, Science and Technology of Japan.

*These authors contributed equally to this work.

†snoda@kuee.kyoto-u.ac.jp

# Supplementary material for

# "Photonic Quantum Computation with Waveguide-Linked Optical Cavities and Quantum Dots"

Makoto Yamaguchi, Takashi Asano, Yoshiya Sato and Susumu Noda

*Department of Electronic Science and Engineering, Kyoto University,*

*Katsura, Nishikyo-ku, Kyoto 615-8510, Japan*

This supplementary material contains:

i) The quantization of the coupled mode theory in section SM-1.

ii) The derivation of the effective Hamiltonian in section SM-2.

iii) The effective Hamiltonian and the specific phase differences for the two-qubit gate operation in section SM-3.

iv) Application of our system as an entangled photon source in section SM-4.

v) One possible structure for many qubits in section SM-5.

## SM-1

In this section, we describe the quantization of the coupled mode theory (CMT) for the system shown in Fig. 1. According to CMT [S1-S3], the amplitude of the $l$-th cavity ($\equiv a_{cl}$) follows a simultaneous differential equation written as

$$\frac{da_{cl}}{dt} = -[i\omega_{cl} + |\kappa_l|^2]a_{cl}(t) + \kappa_l[S^R_{l,\text{in}}(t) + S^L_{l,\text{in}}(t)], \quad (l = 1, 2) \tag{S1}$$

where $S^i_{l,\text{in}}(t)$ and $S^i_{l,\text{out}}(t)$ ($i$ = L, R) are the amplitudes of the incoming and outgoing waves, respectively (Fig. S1) and $\kappa_l$ denotes the coupling coefficient between the $l$-th cavity and the propagating mode in the waveguide, which satisfies the relationship of

$$|\kappa_l|^2 = \Gamma_l. \tag{S2}$$

The squared value of $|a_{cl}(t)|^2$ is the energy in the $l$-th cavity mode and $|S^i_{l,\text{in}}(t)|^2$ and $|S^i_{l,\text{out}}(t)|^2$ express the power propagating in the waveguide. Here, the input-output relations between the incoming and outgoing waves are given by

$$S^i_{l,\text{out}}(t) = S^i_{l,\text{in}}(t) - \kappa^*_l a_{cl}(t), \quad (i = \text{L, R and } l = 1, 2) \tag{S3}$$

and the boundary conditions for the propagation and reflection of light can be written as

$$S^L_{2,\text{in}}(t) = S^L_{1,\text{out}}(t - \tau_{12}), \tag{S4}$$

$$S^R_{1,\text{in}}(t) = S^R_{2,\text{out}}(t - \tau_{12}), \tag{S5}$$

$$S^L_{1,\text{in}}(t) = \exp(i\Delta_1) S^R_{1,\text{out}}(t - \tau_{M1}), \tag{S6}$$

$$S^R_{2,\text{in}}(t) = \exp(i\Delta_2) S^L_{2,\text{out}}(t - \tau_{M2}). \tag{S7}$$

Classical analyses can be performed using Eqs. (S1)-(S7), which form the basis of CMT.

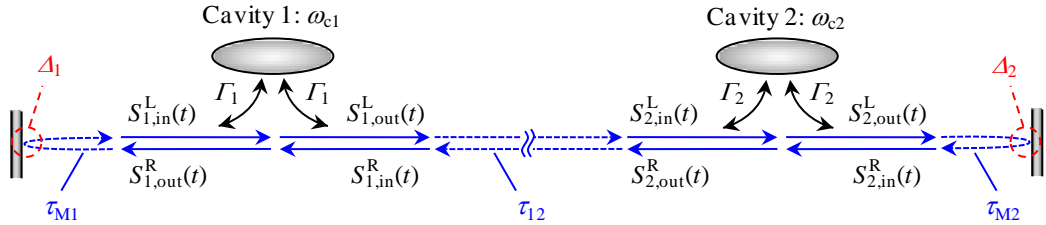

Figure S1. Parameters for CMT of the two cavities evanescently coupled with a waveguide.

In what follows, we show that the Hamiltonian of Eq. (1) is equivalent to the canonical quantization of the CMT described above. For this purpose, it is sufficient to prove that Eqs. (S1)-(S7) can be derived from a *classical* Hamiltonian

$$H = \hbar \sum_{l=1}^{2} \omega_{cl} a^*_{cl} a_{cl} + \hbar \sum_{\lambda \in \{\text{FP}\}} \omega_\lambda a^*_\lambda a_\lambda + i\hbar \sum_{l=1}^{2} \sum_{\lambda \in \{\text{FP}\}} [g_{l,\lambda} a_{cl} a^*_\lambda - g^*_{l,\lambda} a_\lambda a^*_{cl}], \tag{S8}$$

with Poisson brackets of

$$\{a_m, a^*_n\} = -i\hbar^{-1}\delta_{m,n}, \quad \{a_m, a_n\} = 0, \tag{S9}$$

because the canonical quantization of Eq. (S8) with Eq. (S9) directly results in Eq. (1). In our treatment,

the coupling constants are described by the summation of two terms:

$$g_{l,\lambda} = g_{l,\lambda}^{L} + g_{l,\lambda}^{R}, \tag{S10}$$

where the absolute values of $g_{l,\lambda}^{i}$ ($i$ = L, R and $l$ = 1, 2) are defined by

$$\left|g_{l,\lambda}^{L}\right| = \left|g_{l,\lambda}^{R}\right| \equiv \sqrt{\Gamma_{l}/\tau_{P}}, \tag{S11}$$

and the phase differences are determined by

$$\phi_{2}^{L}(\omega_{\lambda}) - \phi_{2}^{R}(\omega_{\lambda}) = \omega_{\lambda}\tau_{M2} + \Delta_{2}, \tag{S12}$$

$$\phi_{1}^{L}(\omega_{\lambda}) - \phi_{2}^{L}(\omega_{\lambda}) = \phi_{2}^{R}(\omega_{\lambda}) - \phi_{1}^{R}(\omega_{\lambda}) = \omega_{\lambda}\tau_{12}, \tag{S13}$$

with the definition

$$\phi_{l}^{i}(\omega_{\lambda}) \equiv \arg[g_{l,\lambda}^{i}]. \tag{S14}$$

The most important point here is that there are phase differences between $g_{l,\lambda}^{L}$ and $g_{l,\lambda}^{R}$ in our treatment (Eqs. (S12) and (S13)), which will introduce consistent input-output relations and boundary conditions as described below.

We will now discuss the canonical equations of motion derived from Eqs. (S8)-(S14), which can be expressed as

$$\frac{dA_{cl}}{dt} = F_{l}^{L}(t) + F_{l}^{R}(t), \qquad (l = 1, 2) \tag{S15}$$

$$\frac{dA_{\lambda}}{dt} = \sum_{j=L,R}\sum_{m=1}^{2} g_{m,\lambda}^{j} A_{cm} \exp(-i[\omega_{cm} - \omega_{\lambda}]t), \qquad (\lambda \in \{\text{FP modes}\}) \tag{S16}$$

with

$$F_{l}^{i}(t) \equiv -\sum_{\lambda \in \{\text{FP}\}} g_{l,\lambda}^{i*} A_{\lambda}(t) \exp(i[\omega_{cl} - \omega_{\lambda}]t), \tag{S17}$$

where $A_{cl}(t)$ and $A_{\lambda}(t)$ are defined as $A_{cl}(t) \equiv a_{cl}(t)\exp(i\omega_{cl}t)$ and $A_{\lambda}(t) \equiv a_{\lambda}(t)\exp(i\omega_{\lambda}t)$, respectively. When Eq. (S16) is formally integrated from $t_0$ to $t$,

$$A_{\lambda}(t) = A_{\lambda}(t_{0}) + \sum_{j=L,R}\sum_{l=1}^{2} g_{l,\lambda}^{j} \int_{t_0}^{t} dt A_{cl}(t') \exp(-i[\omega_{cl} - \omega_{\lambda}]t'), \quad (t > t_0) \tag{S18}$$

can be obtained. Therefore, $F_{1}^{L}$ in Eq. (S15) can be expressed as

$$F_1^L(t) = -\sqrt{\frac{\Gamma_1}{\tau_P}} \sum_{\lambda \in \{FP\}} A_\lambda(t_0) \exp(i[\omega_{c1} - \omega_\lambda]t - i\phi_1^L(\omega_\lambda))$$

$$-\frac{\Gamma_1}{\tau_P} \int_{t_0}^{t} dt A_{c1}(t') \sum_{\lambda \in \{FP\}} \exp(i[\omega_{c1} - \omega_\lambda]t - i[\omega_{c1} - \omega_\lambda]t')$$

$$-\frac{\Gamma_1}{\tau_P} \int_{t_0}^{t} dt A_{c1}(t') \sum_{\lambda \in \{FP\}} \exp(i[\omega_{c1} - \omega_\lambda]t - i[\omega_{c1} - \omega_\lambda]t' - i\omega_\lambda[\tau_{M2} + 2\tau_{12}] - i\Delta_2)$$

$$-\frac{\sqrt{\Gamma_1 \Gamma_2}}{\tau_P} \int_{t_0}^{t} dt A_{c2}(t') \sum_{\lambda \in \{FP\}} \exp(i[\omega_{c1} - \omega_\lambda]t - i[\omega_{c2} - \omega_\lambda]t' - i\omega_\lambda \tau_{12})$$

$$-\frac{\sqrt{\Gamma_1 \Gamma_2}}{\tau_P} \int_{t_0}^{t} dt A_{c2}(t') \sum_{\lambda \in \{FP\}} \exp(i[\omega_{c1} - \omega_\lambda]t - i[\omega_{c2} - \omega_\lambda]t' - i\omega_\lambda[\tau_{M2} + \tau_{12}] - i\Delta_2), \quad (S19)$$

where Eqs. (S11)-(S14) and Eq. (S17) are used in the derivation. Here, it should be noted that the $\lambda$-th FP mode is formed at the frequency of $\omega_\lambda = \lambda \cdot \omega_P - \Delta/\tau_P$ ($\omega_P \equiv 2\pi/\tau_P$, $\Delta \equiv \Delta_1 + \Delta_2$). Therefore, the summation for an arbitrary function $G(\omega_\lambda)$ can be written as

$$\sum_{\lambda \in \{FP\}} G(\omega_\lambda) = \int_0^\infty d\omega G(\omega) D_{FP}(\omega), \quad (S20)$$

where the density of states for the FP modes is

$$D_{FP}(\omega) \equiv \sum_{\lambda=-\infty}^{\infty} \delta(\omega - \lambda\omega_P + \Delta/\tau_P) = \frac{\tau_P}{2\pi} \sum_{\lambda=-\infty}^{\infty} \exp(-i\lambda(\tau_P \omega + \Delta)), \quad (S21)$$

and the right-hand term of Eq. (S21) can be proven to be the Fourier series expansion of the left-hand term. Using these relations with

$$\frac{1}{2\pi} \int_0^\infty d\omega \exp[i(\omega_x - \omega)(t - t')] \cong \delta(t - t'), \quad (S22)$$

Eq. (S19) can be written in the form

$$F_1^L(t) = \kappa_1 S_{1,in}^L(t) \exp(i\omega_{c1}t) - \frac{|\kappa_1|^2}{2} A_{c1}(t), \quad (S23)$$

where $\kappa_l$ ($l = 1, 2$) and $S_{1,in}^i(t)$ are respectively defined as

$$\kappa_l \equiv \sqrt{\Gamma_l} \exp(-i\phi_0), \quad (S24)$$

and

$$S_{1,\text{in}}^{L}(t) \equiv -\frac{1}{\sqrt{\tau_P}} \sum_{\lambda \in \{FP\}} A_\lambda(t_0) \exp(-i\omega_\lambda t - i\phi_1^L(\omega_\lambda) + i\phi_0)$$

$$-\sqrt{\Gamma_1} \sum_{\lambda=1}^{\infty} a_{c1}(t - \lambda\tau_P) u(t - t_0 - \lambda\tau_P) \exp(i\lambda\Delta + i\phi_0)$$

$$-\sqrt{\Gamma_1} \sum_{\lambda=1}^{\infty} a_{c1}(t - \lambda\tau_P + 2\tau_{12} + \tau_{M2}) u(t - t_0 - \lambda\tau_P + 2\tau_{12} + \tau_{M2}) \exp(i\lambda\Delta - i\Delta_2 + i\phi_0)$$

$$-\sqrt{\Gamma_2} \sum_{\lambda=1}^{\infty} a_{c2}(t - \lambda\tau_P + \tau_{12} + \tau_{M2}) u(t - t_0 - \lambda\tau_P + \tau_{12} + \tau_{M2}) \exp(i\lambda\Delta - i\Delta_2 + i\phi_0)$$

$$-\sqrt{\Gamma_2} \sum_{\lambda=1}^{\infty} a_{c2}(t - \lambda\tau_P + \tau_{12}) u(t - t_0 - \lambda\tau_P + \tau_{12}) \exp(i\lambda\Delta + i\phi_0). \tag{S25}$$

Here, $\phi_0$ is an arbitrary phase and $u(t)$ is a step function. Similar analyses can also be performed for $F_1^R$, $F_2^L$ and $F_2^R$ in Eq. (S15), which gives the following general expression:

$$F_l^i(t) = \kappa_l S_{l,\text{in}}^i(t) \exp(i\omega_{cl} t) - \frac{|\kappa_l|^2}{2} A_{cl}(t), \quad (i = L, R \text{ and } l = 1, 2) \tag{S26}$$

with definitions of

$$S_{1,\text{in}}^{R}(t) \equiv -\frac{1}{\sqrt{\tau_P}} \sum_{\lambda \in \{FP\}} A_\lambda(t_0) \exp(-i\omega_\lambda t - i\phi_1^R(\omega_\lambda) + i\phi_0)$$

$$-\sqrt{\Gamma_1} \sum_{\lambda=1}^{\infty} a_{c1}(t - \lambda\tau_P) u(t - t_0 - \lambda\tau_P) \exp(i\lambda\Delta + i\phi_0)$$

$$-\sqrt{\Gamma_1} \sum_{\lambda=1}^{\infty} a_{c1}(t - \lambda\tau_P + \tau_{M1}) u(t - t_0 - \lambda\tau_P + \tau_{R1}) \exp(i\lambda\Delta - i\Delta_1 + i\phi_0)$$

$$-\sqrt{\Gamma_2} \sum_{\lambda=1}^{\infty} a_{c2}(t - [\lambda-1]\tau_P - \tau_{12}) u(t - t_0 - [\lambda-1]\tau_P - \tau_{12}) \exp(i[\lambda-1]\Delta + i\phi_0)$$

$$-\sqrt{\Gamma_2} \sum_{\lambda=1}^{\infty} a_{c2}(t - \lambda\tau_P + \tau_{M1} + \tau_{12}) u(t - t_0 - \lambda\tau_P + \tau_{M1} + \tau_{12}) \exp(i\lambda\Delta - i\Delta_1 + i\phi_0), \tag{S27}$$

$$S_{2,\text{in}}^{L}(t) \equiv -\frac{1}{\sqrt{\tau_P}} \sum_{\lambda \in \{FP\}} A_\lambda(t_0) \exp(-i\omega_\lambda t - i\phi_2^L(\omega_\lambda) + i\phi_0)$$

$$-\sqrt{\Gamma_2} \sum_{\lambda=1}^{\infty} a_{c2}(t - \lambda\tau_P) u(t - t_0 - \lambda\tau_P) \exp(i\lambda\Delta + i\phi_0)$$

$$-\sqrt{\Gamma_2} \sum_{\lambda=1}^{\infty} a_{c2}(t - \lambda\tau_P + \tau_{M2}) u(t - t_0 - \lambda\tau_P + \tau_{M2}) \exp(i\lambda\Delta - i\Delta_2 + i\phi_0)$$

$$-\sqrt{\Gamma_1} \sum_{\lambda=1}^{\infty} a_{c1}(t - \lambda\tau_P + \tau_{M2} + \tau_{12}) u(t - t_0 - \lambda\tau_P + \tau_{M2} + \tau_{12}) \exp(i\lambda\Delta - i\Delta_2 + i\phi_0)$$

$$-\sqrt{\Gamma_1}\sum_{\lambda=0}^{\infty}a_{c1}(t-\lambda\tau_P-\tau_{12})u(t-t_0-\lambda\tau_P-\tau_{12})\exp(i\lambda\Delta+i\phi_0), \tag{S28}$$

$$S_{2,\text{in}}^{R}(t) \equiv -\frac{1}{\sqrt{\tau_P}}\sum_{\lambda\in\{FP\}}A_\lambda(t_0)\exp(-i\omega_\lambda t-i\phi_2^R(\omega_\lambda)+i\phi_0)$$

$$-\sqrt{\Gamma_2}\sum_{\lambda=1}^{\infty}a_{c2}(t-\lambda\tau_P)u(t-t_0-\lambda\tau_P)\exp(i\lambda\Delta+i\phi_0)$$

$$-\sqrt{\Gamma_2}\sum_{\lambda=0}^{\infty}a_{c2}(t-\lambda\tau_P-\tau_{M2})u(t-t_0-\lambda\tau_P-\tau_{M2})\exp(i\lambda\Delta+i\Delta_2+i\phi_0)$$

$$-\sqrt{\Gamma_1}\sum_{\lambda=1}^{\infty}a_{c1}(t-\lambda\tau_P+\tau_{12})u(t-t_0-\lambda\tau_P+\tau_{12})\exp(i\lambda\Delta+i\phi_0)$$

$$-\sqrt{\Gamma_1}\sum_{\lambda=0}^{\infty}a_{c1}(t-\lambda\tau_P-\tau_{12}-\tau_{M2})u(t-t_0-\lambda\tau_P-\tau_{12}-\tau_{M2})\exp(i\lambda\Delta+i\Delta_2+i\phi_0). \tag{S29}$$

Therefore, Eq. (S26) can be substituted into Eq. (S15), yielding

$$\frac{dA_{cl}}{dt} = -|\kappa_l|^2 A_{cl}(t) + \sum_{i=L,R}\kappa_l S_{l,\text{in}}^i(t)\exp(i\omega_{cl}t), \quad (l=1,2) \tag{S30}$$

which is equivalent to Eq. (S1) in CMT with the definition of $A_{cl}(t) \equiv a_{cl}(t)\exp(i\omega_{cl}t)$. Furthermore, one can find from Eq. (S25) and Eqs. (S27)-(S29) that the following relationships hold:

$$S_{1,\text{in}}^{L}(t) = \exp(i\Delta_1)[S_{1,\text{in}}^{R}(t-\tau_{M1}) - \kappa_1^* a_{c1}(t-\tau_{M1})], \tag{S31}$$

$$S_{2,\text{in}}^{R}(t) = \exp(i\Delta_2)[S_{2,\text{in}}^{L}(t-\tau_{M2}) - \kappa_2^* a_{c2}(t-\tau_{M2})], \tag{S32}$$

$$S_{2,\text{in}}^{L}(t) = S_{1,\text{in}}^{L}(t-\tau_{12}) - \kappa_1^* a_{c1}(t-\tau_{12}), \tag{S33}$$

$$S_{1,\text{in}}^{R}(t) = S_{2,\text{in}}^{R}(t-\tau_{12}) - \kappa_2^* a_{c2}(t-\tau_{12}), \tag{S34}$$

when $t_0 = -\infty$. Even though these relationships partly correspond to Eqs. (S3)-(S7) in CMT, the outgoing waves are not included. Therefore, the outgoing waves must be introduced separately, which can be achieved by using

$$A_\lambda(t) = A_\lambda(t_1) - \sum_{j=L,R}\sum_{l=1}^{2}g_{l,\lambda}^j\int_{t}^{t_1}dt A_{cl}(t')\exp(-i[\omega_{cl}-\omega_\lambda]t'), \quad (t<t_1) \tag{S35}$$

instead of Eq. (S18) in the analyses. In this case, the evolution of $A_{cl}(t)$ is written as if the time flows backward from a point in the future ($t_1$). As a result, the incoming waves and the negative sign related to the loss in Eq. (S26) are replaced by the outgoing waves and a positive sign, respectively (this treatment is closely analogous to the quantum Langevin equations [S4]). Thus, $F_l^i(t)$ can also be described with the outgoing waves:

$$F_l^i(t) = \kappa_l S_{l,\text{out}}^i(t)\exp(i\omega_{cl}t) + \frac{|\kappa_l|^2}{2}A_{cl}(t), \quad (i = \text{L, R and } l = 1, 2) \tag{S36}$$

Therefore, the relationship between the incoming and outgoing waves is obtained as

$$S_{l,\text{out}}^i(t) = S_{l,\text{in}}^i(t) - \kappa_l^* a_{cl}(t), \quad (i = \text{L, R and } l = 1, 2) \tag{S37}$$

by comparing Eqs. (S26) and (S36), which is the same as Eq. (S3) in CMT. Furthermore, Eqs. (S31)-(S34) are also found to be the same as Eqs. (S4)-(S7) by using Eq. (S37). Thus, all of equations in CMT (Eqs. (S1)-(S7)) can be derived from Eqs. (S8)-(S14) with the density of states for the FP modes (Eq. (S21)). Consequently, the canonical quantization of Eq. (S8) with Eq. (S9) directly results in Eq. (1). We have thus demonstrated the quantization of CMT for the system shown in Fig. 1.

## SM-2

Here, we perform the derivation of the effective Hamiltonian described by Eqs. (2)-(5) where the non-Markov quantum master equation (QME) is adopted for the elimination of the waveguide's degrees of freedom. First, we briefly review the non-Markov QME for a Hamiltonian $\hat{H}_\text{tot} = \hat{H}_\text{S} + \hat{H}_\text{R} + \hat{H}_\text{SR}$, where $\hat{H}_\text{S}$ and $\hat{H}_\text{R}$ denote the Hamiltonian of a system and a reservoir, respectively, and $\hat{H}_\text{SR}$ denotes the interactions between them. In general, $\hat{H}_\text{SR}$ can be described in the form

$$\hat{H}_\text{SR}^\text{I}(t) = \sum_\alpha \hat{S}_\alpha^\text{I}(t) \otimes \hat{R}_\alpha^\text{I}(t), \tag{S38}$$

in the interaction picture (expressed by the superscript "I"). Here, $\hat{S}_\alpha^\text{I}(t)$ and $\hat{R}_\alpha^\text{I}(t)$ are the operators of the system and the reservoir, respectively. When the density operator of the whole system ($\hat{\rho}_\text{SR}^\text{I}(t)$) can be described within the Born approximation

$$\hat{\rho}_\text{SR}^\text{I}(t) \cong \hat{\rho}_\text{S}^\text{I}(t) \otimes \hat{\rho}_{\text{R},0}, \tag{S39}$$

the non-Markov QME for the reduced density operator of the system ($\hat{\rho}_\text{S}^\text{I}(t)$) can be expressed as

$$\frac{d}{dt}\hat{\rho}_S^I(t) = \sum_{\alpha,\beta} \left.\frac{d\hat{\rho}_S^I(t)}{dt}\right|_{\alpha\beta}, \tag{S40}$$

$$\left.\frac{d\hat{\rho}_S^I(t)}{dt}\right|_{\alpha\beta} \equiv -\hbar^{-2}\int_0^t dt' C_{\alpha\beta}(t')[\hat{S}_\alpha^I(t)\hat{S}_\beta^I(t-t')\hat{\rho}_S^I(t-t') - \hat{S}_\beta^I(t-t')\hat{\rho}_S^I(t-t')\hat{S}_\alpha^I(t)]$$

$$-\hbar^{-2}\int_0^t dt' C_{\alpha\beta}(-t')[\hat{\rho}_S^I(t-t')\hat{S}_\alpha^I(t-t')\hat{S}_\beta^I(t) - \hat{S}_\beta^I(t)\hat{\rho}_S^I(t-t')\hat{S}_\alpha^I(t-t')], \tag{S41}$$

where $C_{\alpha\beta}(t)$ is the correlation function of the reservoir, written as

$$C_{\alpha\beta}(t) \equiv \text{Tr}_R[\hat{R}_\alpha^I(t)\hat{R}_\beta^I(0)\hat{\rho}_{R,0}]. \tag{S42}$$

Here, $\hat{\rho}_{R,0}$ denotes the initial density operator of the reservoir. In general, the reservoir is sufficiently large and its density of states is taken as a quasi-continuum, which allows the use of the Born approximation.

Now, the general description of the non-Markov QME is applied to the Hamiltonian described by Eq. (1) with Eqs. (S10)-(S14) where the first and second cavities are the system, whereas the waveguide is taken as the reservoir in our treatment. The waveguide in our system is not sufficiently large to be treated as a quasi-continuum, which would invalidate the Born approximation. Therefore, we have to assume alternatively that $\omega_{cl}$ is sufficiently detuned from the FP modes in the waveguide:

$$|\omega_{cl} - \omega_\lambda| \gg |g_{l,\lambda}|. \quad (l = 1, 2) \tag{S43}$$

In this case, the change of FP modes from their initial states would be negligible, allowing the use of the Born approximation again, which is schematically shown in Fig. S2.

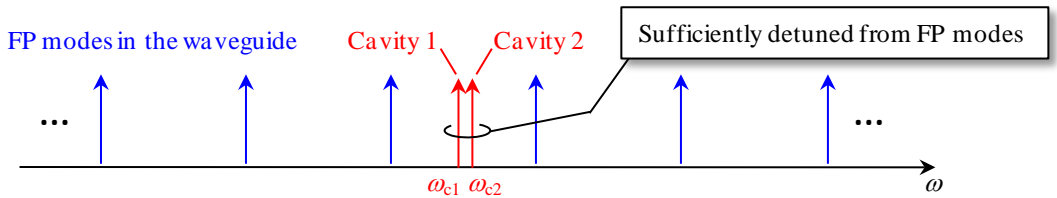

Figure S2. Schematic picture of the situation for the Born approximation in our system.

In this limit, the non-Markov QME can be applied to our system (Eq. (1)) and the operators in Eq. (S38) can be defined as

$$\hat{S}_1^I(t) \equiv \hat{a}_{c1} \exp(-i\omega_{c1}t), \quad \hat{R}_1^I(t) \equiv i\hbar \sum_{\lambda \in \{FP\}} g_{1,\lambda} \hat{a}_\lambda^\dagger \exp(i\omega_\lambda t), \tag{S44}$$

$$\hat{S}_2^I(t) \equiv \hat{a}_{c1}^\dagger \exp(i\omega_{c1}t), \quad \hat{R}_2^I(t) \equiv -i\hbar \sum_{\lambda \in \{FP\}} g_{1,\lambda}^* \hat{a}_\lambda \exp(-i\omega_\lambda t), \tag{S45}$$

$$\hat{S}_3^I(t) \equiv \hat{a}_{c2} \exp(-i\omega_{c2}t), \quad \hat{R}_3^I(t) \equiv i\hbar \sum_{\lambda \in \{FP\}} g_{2,\lambda} \hat{a}_\lambda^\dagger \exp(i\omega_\lambda t), \tag{S46}$$

$$\hat{S}_4^I(t) \equiv \hat{a}_{c2}^\dagger \exp(i\omega_{c2}t), \quad \hat{R}_4^I(t) \equiv -i\hbar \sum_{\lambda \in \{FP\}} g_{2,\lambda}^* \hat{a}_\lambda \exp(-i\omega_\lambda t), \tag{S47}$$

with definitions of

$$\hat{H}_S = \hbar \sum_{l=1}^{2} \omega_{cl} \hat{a}_{cl}^\dagger \hat{a}_{cl}, \quad \hat{H}_R = \hbar \sum_{\lambda \in \{FP\}} \omega_\lambda \hat{a}_\lambda^\dagger \hat{a}_\lambda. \tag{S48}$$

In this case, one can find that $C_{\alpha\beta}(t)$ is non-zero only for a set of

$$(\alpha, \beta) = (2, 1), (4, 1), (2, 3), (4, 3), \tag{S49}$$

when assuming that the photon temperature of the waveguide is zero:

$$\langle \hat{a}_i^\dagger \hat{a}_j^\dagger \rangle_R = \langle \hat{a}_i \hat{a}_j \rangle_R = \langle \hat{a}_i^\dagger \hat{a}_j \rangle_R = 0, \quad \langle \hat{a}_i \hat{a}_j^\dagger \rangle_R = \delta_{i,j}, \quad (i, j \in \{FP \text{ modes}\}) \tag{S50}$$

where $\langle \ldots \rangle_R \equiv \text{Tr}_R[\ldots \hat{\rho}_{R,0}]$ and $\delta_{i,j}$ is the Kronecker's delta. Therefore, in the case of $(\alpha, \beta) = (2, 1)$, the correlation function of $C_{21}(t)$ can be calculated from Eqs. (S42), (S44), and (S45) as

$$C_{21}(t) = 2\hbar^2 \Gamma_1 \sum_{\lambda=-\infty}^{\infty} \delta(t + \lambda\tau_P) \exp(-i\lambda\Delta)$$

$$+ \hbar^2 \Gamma_1 \sum_{\lambda=-\infty}^{\infty} \delta(t + \lambda\tau_P + \tau_{M2} + 2\tau_{12}) \exp(-i\lambda\Delta - i\Delta_2)$$

$$+ \hbar^2 \Gamma_1 \sum_{\lambda=-\infty}^{\infty} \delta(t + \lambda\tau_P - \tau_{M2} - 2\tau_{12}) \exp(-i\lambda\Delta + i\Delta_2), \tag{S51}$$

where Eqs. (S10)-(S14) and (S20)-(S22) are used in the derivation. Therefore, by substituting Eq. (S51) into Eq. (S41), $d\hat{\rho}_S^I(t)/dt|_{21}$ can be obtained as

$$\left.\frac{d\hat{\rho}_S^I(t)}{dt}\right|_{21} = -\Gamma_1 [\hat{S}_2^I(t), \hat{S}_1^I \hat{\rho}_S^I u(t)] - 2\Gamma_1 \sum_{\lambda=1}^{\infty} \exp(i\lambda\Delta)[\hat{S}_2^I(t), \hat{S}_1^I \hat{\rho}_S^I u(t - \lambda\tau_P)]$$

$$- \Gamma_1 \sum_{\lambda=1}^{\infty} \exp(i\lambda\Delta - i\Delta_2)[\hat{S}_2^I(t), \hat{S}_1^I \hat{\rho}_S^I u(t - \lambda\tau_P + \tau_{M2} + 2\tau_{12})]$$

$$-\Gamma_1 \sum_{\lambda=0}^{\infty} \exp(i\lambda\Delta + i\Delta_2)[\hat{S}_2^I(t), \hat{S}_1^I \hat{\rho}_S^I u(t - \lambda\tau_P - \tau_{M2} - 2\tau_{21})]$$

$$-\Gamma_1[\hat{\rho}_S^I \hat{S}_2^I u(t), \hat{S}_1^I(t)] - 2\Gamma_1 \sum_{\lambda=1}^{\infty} \exp(-i\lambda\Delta)[\hat{\rho}_S^I \hat{S}_2^I u(t - \lambda\tau_P), \hat{S}_1^I(t)]$$

$$-\Gamma_1 \sum_{\lambda=0}^{\infty} \exp(-i\lambda\Delta - i\Delta_2)[\hat{\rho}_S^I \hat{S}_2^I u(t - \lambda\tau_P - \tau_{M2} - 2\tau_{21}), \hat{S}_1^I(t)]$$

$$-\Gamma_1 \sum_{\lambda=1}^{\infty} \exp(-i\lambda\Delta + i\Delta_2)[\hat{\rho}_S^I \hat{S}_2^I u(t - \lambda\tau_P + \tau_{M2} + 2\tau_{21}), \hat{S}_1^I(t)], \tag{S52}$$

where $\hat{A}^I(t)\hat{B}^I(t)u(t)$ is expressed as $\hat{A}^I\hat{B}^I u(t)$ to simplify the description ($\hat{A}^I(t)$ and $\hat{B}^I(t)$ are arbitrary operators). However, this is so complex that Eq. (S52) should be approximated further. Therefore, the following approximation is introduced

$$\hat{\rho}_S^I(t')u(t') \cong \hat{\rho}_S^I(t)u(t), \qquad (t \leq t' < t + \tau_P) \tag{S53}$$

with the assumption of $\tau_P \ll 1/\Gamma_1, 1/\Gamma_2$. Taking this approximation together with Eqs. (S44) and (S45), Eq. (S52) can be simplified as

$$\left.\frac{d\hat{\rho}_S^I(t)}{dt}\right|_{21} = -\Gamma_1[\hat{S}_2^I(t), \hat{S}_1^I \hat{\rho}_S^I u(t)] - \Gamma_1[\hat{\rho}_S^I \hat{S}_2^I u(t), \hat{S}_1^I(t)]$$

$$-\Gamma_1[2 + \exp[-i\theta_{M1}(\omega_{c1})] + \exp[-i2\theta_{12}(\omega_{c1}) - i\theta_{M2}(\omega_{c1})]] \cdot [\hat{S}_2^I(t), \hat{X}(t)]$$

$$-\Gamma_1[2 + \exp[i\theta_{M1}(\omega_{c1})] + \exp[2i\theta_{12}(\omega_{c1}) + i\theta_{M2}(\omega_{c1})]] \cdot [\hat{Y}(t), \hat{S}_1^I(t)], \tag{S54}$$

where $\hat{X}(t)$ and $\hat{Y}(t)$ are defined as

$$\hat{X}(t) \equiv \sum_{\lambda=1}^{\infty} \exp(i\lambda\Delta)\hat{S}_1^I \hat{\rho}_S^I u(t - \lambda\tau_P), \quad \hat{Y}(t) \equiv \sum_{\lambda=1}^{\infty} \exp(-i\lambda\Delta)\hat{\rho}_S^I \hat{S}_2^I u(t - \lambda\tau_P), \tag{S55}$$

respectively. $\theta_{M1}(\omega)$, $\theta_{M2}(\omega)$, and $\theta_{12}(\omega)$ are the phase differences caused by the propagation of light, described by

$$\theta_{M1}(\omega) \equiv \omega\tau_{M1} + \Delta_1, \quad \theta_{M2}(\omega) \equiv \omega\tau_{M2} + \Delta_2, \quad \theta_{12}(\omega) \equiv \omega\tau_{12}. \tag{S56}$$

Here, one can find that Eq. (S55) satisfies the following relationships:

$$\hat{X}(t + \tau_P) = \exp(i\Delta)[\hat{S}_1^I \hat{\rho}_S^I u(t) + \hat{X}(t)], \quad \hat{Y}(t + \tau_P) = \exp(-i\Delta)[\hat{\rho}_S^I \hat{S}_2^I u(t) + \hat{Y}(t)], \tag{S57}$$

and $\hat{X}(t + \tau_P)$ and $\hat{Y}(t + \tau_P)$ can be approximated by

$$\hat{X}(t + \tau_P) \cong \exp(-i\omega_{c1}\tau_P)\hat{X}(t), \quad \hat{Y}(t + \tau_P) \cong \exp(i\omega_{c1}\tau_P)\hat{Y}(t), \tag{S58}$$

through Eq. (S53) together with Eqs. (S44) and (S45). Therefore, $\hat{X}(t)$ and $\hat{Y}(t)$ can be approximated as

$$\hat{X}(t) \cong \frac{1}{\exp[-i\theta_P(\omega_{c1})]-1} \hat{S}_1^I \hat{\rho}_S^I u(t), \quad \hat{Y}(t) \cong \frac{1}{\exp[i\theta_P(\omega_{c1})]-1} \hat{\rho}_S^I \hat{S}_2^I u(t), \tag{S59}$$

where $\theta_P(\omega) \equiv \theta_{M1}(\omega) + \theta_{M2}(\omega) + 2\theta_{12}(\omega)$ $(= \omega\tau_P + \Delta)$ is the phase difference for the round trip of light along the waveguide. It should be noted that $\theta_P(\omega_\lambda) = \lambda \cdot 2\pi$ and $\exp[\pm i\theta_P(\omega_\lambda)] = 1$ are obtained for the $\lambda$-th FP modes because $\omega_\lambda = \lambda \cdot \omega_P - \Delta/\tau_P$, as described in section SM-1. On the other hand, $\exp[\pm i\theta_P(\omega_{cl})] \neq 1$ ($l = 1, 2$) because $\omega_{cl}$ is sufficiently detuned from the FP modes, in order to apply the Born approximation. Therefore, Eqs. (S54) and (S59) yield

$$\left.\frac{d\hat{\rho}_S^I(t)}{dt}\right|_{21} = -i\chi(\theta_{M1}(\omega_{c1}), \theta_{M2}(\omega_{c1}) + 2\theta_{12}(\omega_{c1}), \theta_P(\omega_{c1}))\Gamma_1[\hat{S}_2^I\hat{S}_1^I(t), \hat{\rho}_S^I u(t)], \tag{S60}$$

where the function $\chi(x, y, z)$ is defined as $\chi(x, y, z) \equiv [\cos([x+y]/2)+\cos([x-y]/2)]/\sin(z/2)$. Thus, a simple description is obtained for $d\hat{\rho}_S^I(t)/dt|_{21}$. Similar analyses are also possible for the remaining set of $(\alpha, \beta) = (4, 1), (2, 3), (4, 3)$ in Eq. (S49) and the following equations can be obtained:

$$\left.\frac{d\hat{\rho}_S^I(t)}{dt}\right|_{41} = -i\chi(\theta_{M1}(\omega_{c1}), \theta_{M2}(\omega_{c1}), \theta_P(\omega_{c1}))\sqrt{\Gamma_1\Gamma_2}[\hat{S}_4^I(t), \hat{S}_1^I\hat{\rho}_S^I u(t)]$$

$$+ i\chi(\theta_{M1}(\omega_{c2}), \theta_{M2}(\omega_{c2}), \theta_P(\omega_{c2}))\sqrt{\Gamma_1\Gamma_2}[\hat{\rho}_S^I\hat{S}_4^I u(t), \hat{S}_1^I(t)], \tag{S61}$$

$$\left.\frac{d\hat{\rho}_S^I(t)}{dt}\right|_{23} = -i\chi(\theta_{M1}(\omega_{c2}), \theta_{M2}(\omega_{c2}), \theta_P(\omega_{c2}))\sqrt{\Gamma_1\Gamma_2}[\hat{S}_2^I(t), \hat{S}_3^I\hat{\rho}_S^I u(t)]$$

$$+ i\chi(\theta_{M1}(\omega_{c1}), \theta_{M2}(\omega_{c1}), \theta_P(\omega_{c1}))\sqrt{\Gamma_1\Gamma_2}[\hat{\rho}_S^I\hat{S}_2^I u(t), \hat{S}_3^I(t)], \tag{S62}$$

$$\left.\frac{d\hat{\rho}_S^I(t)}{dt}\right|_{43} = -i\chi(\theta_{M2}(\omega_{c2}), \theta_{M1}(\omega_{c2}) + 2\theta_{12}(\omega_{c2}), \theta_P(\omega_{c2}))\Gamma_2[\hat{S}_4^I\hat{S}_3^I(t), \hat{\rho}_S^I u(t)]. \tag{S63}$$

Therefore, Eq. (S40) can be written as

$$\frac{d}{dt}\hat{\rho}_S^I(t) = -i\chi(\theta_{M1}(\omega_{c1}), \theta_{M2}(\omega_{c1}) + 2\theta_{12}(\omega_{c1}), \theta_P(\omega_{c1}))\Gamma_1[\hat{S}_2^I\hat{S}_1^I(t), \hat{\rho}_S^I(t)]$$

$$- i\chi(\theta_{M2}(\omega_{c2}), \theta_{M1}(\omega_{c2}) + 2\theta_{12}(\omega_{c2}), \theta_P(\omega_{c2}))\Gamma_2[\hat{S}_4^I\hat{S}_3^I(t), \hat{\rho}_S^I(t)]$$

$$- i\chi(\theta_{M1}(\omega_{c1}), \theta_{M2}(\omega_{c1}), \theta_P(\omega_{c1}))\sqrt{\Gamma_1\Gamma_2}[[\hat{S}_4^I(t), \hat{S}_1^I\hat{\rho}_S^I(t)] - [\hat{\rho}_S^I\hat{S}_2^I(t), \hat{S}_3^I(t)]]$$

$$-\mathrm{i}\chi(\theta_{\mathrm{M1}}(\omega_{c2}),\theta_{\mathrm{M2}}(\omega_{c2}),\theta_{\mathrm{P}}(\omega_{c2}))\sqrt{\Gamma_1\Gamma_2}[[\hat{S}_2^{\mathrm{I}}(t),\hat{S}_3^{\mathrm{I}}\hat{\rho}_{\mathrm{S}}^{\mathrm{I}}(t)]-[\hat{\rho}_{\mathrm{S}}^{\mathrm{I}}\hat{S}_4^{\mathrm{I}}(t),\hat{S}_1^{\mathrm{I}}(t)]],\tag{S64}$$

by substituting Eqs. (S60)-(S63) into Eq. (S40). As a result, the following QME can be obtained:

$$\frac{\mathrm{d}}{\mathrm{d}t}\hat{\rho}_{\mathrm{S}}(t) = -\frac{\mathrm{i}}{\hbar}[\hat{H}_{\mathrm{S}},\hat{\rho}_{\mathrm{S}}]$$

$$-\mathrm{i}\chi(\theta_{\mathrm{M1}}(\omega_{c1}),\theta_{\mathrm{M2}}(\omega_{c1})+2\theta_{12}(\omega_{c1}),\theta_{\mathrm{P}}(\omega_{c1}))\Gamma_1[\hat{a}_{c1}^{\dagger}\hat{a}_{c1},\hat{\rho}_{\mathrm{S}}(t)]$$

$$-\mathrm{i}\chi(\theta_{\mathrm{M2}}(\omega_{c2}),\theta_{\mathrm{M1}}(\omega_{c2})+2\theta_{12}(\omega_{c2}),\theta_{\mathrm{P}}(\omega_{c2}))\Gamma_2[\hat{a}_{c2}^{\dagger}\hat{a}_{c2},\hat{\rho}_{\mathrm{S}}(t)]$$

$$-\mathrm{i}\chi(\theta_{\mathrm{M1}}(\omega_{c1}),\theta_{\mathrm{M2}}(\omega_{c1}),\theta_{\mathrm{P}}(\omega_{c1}))\sqrt{\Gamma_1\Gamma_2}[[\hat{a}_{c2}^{\dagger},\hat{a}_{c1}\hat{\rho}_{\mathrm{S}}(t)]-[\hat{\rho}_{\mathrm{S}}(t)\hat{a}_{c1}^{\dagger},\hat{a}_{c2}]]$$

$$-\mathrm{i}\chi(\theta_{\mathrm{M1}}(\omega_{c2}),\theta_{\mathrm{M2}}(\omega_{c2}),\theta_{\mathrm{P}}(\omega_{c2}))\sqrt{\Gamma_1\Gamma_2}[[\hat{a}_{c1}^{\dagger},\hat{a}_{c2}\hat{\rho}_{\mathrm{S}}(t)]-[\hat{\rho}_{\mathrm{S}}(t)\hat{a}_{c2}^{\dagger},\hat{a}_{c1}]],\tag{S65}$$

when converted into the Schrödinger's picture using Eqs. (S44)-(S48). Therefore, it can easily be found that the evolution of the system can be described by the effective Hamiltonian of Eqs. (2)-(5) when the two cavities have the same properties (i.e., the resonant frequencies $\omega_{c1} = \omega_{c2} \equiv \omega_0$ and the emission rate $\Gamma_1 = \Gamma_2 \equiv \Gamma_c$):

$$\frac{\mathrm{d}}{\mathrm{d}t}\hat{\rho}_{\mathrm{S}}(t) = -\frac{\mathrm{i}}{\hbar}[\hat{H}_{\mathrm{I}}^{\mathrm{eff}},\hat{\rho}_{\mathrm{S}}].\tag{S66}$$

Thus, we have shown that the effective Hamiltonian described by Eqs. (2)-(5) can be derived from Eq. (1) using Eqs. (S10)-(S14). It should be noted that our approach works well because all of the input-output relations and boundary conditions are included in the original Hamiltonian. Therefore, the effect of propagation time can be taken into account (although approximated), which was difficult for the previous approach [S4, S5]. In this sense, the current approach is advantageous.

**SM-3**

In the previous sections (SM-1 and SM-2), the quantization of CMT and the derivation of the effective Hamiltonian have been described for the system shown in Fig. 1 (or Fig. S1). The presented treatment can easily be expanded for more complicated systems. As a result, the effective Hamiltonian $\hat{H}_{\mathrm{II}}^{\mathrm{eff}}$ for the system shown in Fig. 2 can also be derived as

$$\hat{H}_{\mathrm{II}}^{\mathrm{eff}} = \hbar\sum_{l=3}^{6}\omega_{cl}^{\mathrm{eff}}\hat{a}_{cl}^{\dagger}\hat{a}_{cl} + \hbar[\omega_{\mathrm{x}}^{\mathrm{eff}}\hat{\sigma}_{\mathrm{x,x}} + \omega_{\mathrm{y}}^{\mathrm{eff}}\hat{\sigma}_{\mathrm{y,y}}] + \hat{V}^{(\mathrm{i})} + \hat{V}^{(\mathrm{ii})},\tag{S67}$$

with

$$\hat{V}^{(i)} \equiv \hbar g_{34}^{\text{eff}} [\hat{a}_{c3}^\dagger \hat{a}_{c4} + \text{h.c.}] + \hbar g_{y3}^{\text{eff}} [\hat{a}_{c3}^\dagger \hat{\sigma}_{g,y} + \text{h.c.}] + \hbar g_{y4}^{\text{eff}} [\hat{a}_{c4}^\dagger \hat{\sigma}_{g,y} + \text{h.c.}],  \quad\quad (S68)$$

$$\hat{V}^{(ii)} \equiv \hbar g_{56}^{\text{eff}} [\hat{a}_{c5}^\dagger \hat{a}_{c6} + \text{h.c.}] + \hbar g_{x5}^{\text{eff}} [\hat{a}_{c5}^\dagger \hat{\sigma}_{g,x} + \text{h.c.}] + \hbar g_{x6}^{\text{eff}} [\hat{a}_{c6}^\dagger \hat{\sigma}_{g,x} + \text{h.c.}],  \quad\quad (S69)$$

where the definitions of the effective resonant frequencies and the effective coupling constants are summarized in Table SI. Here, $\theta_{P1} \equiv \theta_{M3} + 2\theta_{34} + 2\theta_{4y} + \theta_{My}$ and $\theta_{P2} \equiv \theta_{Mx} + 2\theta_{5x} + 2\theta_{56} + \theta_{M6}$ are the phase differences for the round trip of light along the waveguides 1 and 2, respectively. Therefore, the effective interactions of the system can be controlled by changing the phase differences, allowing the q-CNOT operation. The specific phase differences at each step (A)-(C) in the q-CNOT operation are described in Table SII, from which it can be confirmed that the Hamiltonian described in the main text is realized by using these phase differences.

Table SI. Definitions of the effective resonant frequencies and the effective coupling constants.

| Variable | Definition | Variable | Definition |
|---|---|---|---|
| $\omega_{c3}^{\text{eff}}$ | $\omega_0 + \chi(\theta_{M3}, 2\theta_{34} + 2\theta_{4y} + \theta_{My}, \theta_{P1})\Gamma$ | $g_{34}^{\text{eff}}$ | $\chi(\theta_{M3}, 2\theta_{4y} + \theta_{My}, \theta_{P1})\Gamma$ |
| $\omega_{c4}^{\text{eff}}$ | $\omega_0 + \chi(\theta_{M3} + 2\theta_{34}, 2\theta_{4y} + \theta_{My}, \theta_{P1})\Gamma$ | $g_{y3}^{\text{eff}}$ | $\sqrt{2}\chi(\theta_{M3}, \theta_{My}, \theta_{P1})\Gamma$ |
| $\omega_{c5}^{\text{eff}}$ | $\omega_0 + 2\chi(\theta_{M6} + 2\theta_{56}, 2\theta_{5x} + \theta_{Mx}, \theta_{P2})\Gamma$ | $g_{y4}^{\text{eff}}$ | $\sqrt{2}\chi(\theta_{M3} + 2\theta_{34}, \theta_{My}, \theta_{P1})\Gamma$ |
| $\omega_{c6}^{\text{eff}}$ | $\omega_0 + 2\chi(\theta_{M6}, 2\theta_{56} + 2\theta_{5x} + \theta_{Mx}, \theta_{P2})\Gamma$ | $g_{56}^{\text{eff}}$ | $2\chi(\theta_{M6}, 2\theta_{5x} + \theta_{Mx}, \theta_{P2})\Gamma$ |
| $\omega_{x}^{\text{eff}}$ | $\omega_0 + 2\chi(\theta_{M6} + 2\theta_{56} + 2\theta_{5x}, \theta_{Mx}, \theta_{P2})\Gamma$ | $g_{x5}^{\text{eff}}$ | $2\chi(\theta_{M6} + 2\theta_{56}, \theta_{Mx}, \theta_{P2})\Gamma$ |
| $\omega_{y}^{\text{eff}}$ | $\omega_0 + 2\chi(\theta_{M3} + 2\theta_{34} + 2\theta_{4y}, \theta_{My}, \theta_{P1})\Gamma$ | $g_{x6}^{\text{eff}}$ | $2\chi(\theta_{M6}, \theta_{Mx}, \theta_{P2})\Gamma$ |

Although the q-CNOT gate operation is focused on in the main text as a standard two-qubit gate, it is possible to implement the CNOT gate by adding another step (D) subsequent to step (C), also shown in Table SII. In step (D), the resonant frequency of the sixth cavity is detuned ($\omega_{c6}^{\text{eff}} - \omega_{cl}^{\text{eff}} = 2\Gamma$ for $l = 1, 2,$ …, 5) without any coupling of the elements ($\hat{V}^{(i)} = \hat{V}^{(ii)} = 0$). Therefore, the one-qubit operation $|b\rangle_{L2} \rightarrow$ $(-1)^b |b\rangle_{L2}$ can be achieved with the interaction time $\pi/2\Gamma$, which completes the CNOT operation.

Table SII. Phase differences for the two-qubit gate operation.

| Operation | $\theta_{M3}$ | $\theta_{34}$ | $\theta_{4y}$ | $\theta_{My}$ | $\theta_{Mx}$ | $\theta_{5x}$ | $\theta_{56}$ | $\theta_{M6}$ | Time |
|---|---|---|---|---|---|---|---|---|---|
| step (A) | $\pi$ | $\pi/2$ | 0 | $\pi$ | $\pi/2$ | $\pi$ | $-\pi/4$ | $\pi$ | $\pi/4\Gamma$ |
| step (B) | $\pi/2$ | 0 | $\pi$ | $\pi/2$ | $\pi$ | 0 | $\pi/2$ | $\pi$ | $\pi/2\Gamma$ |
| step (C) | $\pi$ | $\pi/2$ | 0 | $\pi$ | $\pi/2$ | $\pi$ | $-\pi/4$ | $\pi$ | $\pi/4\Gamma$ |
| step (D) | $\pi$ | $\pi/2$ | 0 | $\pi$ | $\pi$ | 0 | $5\pi/4$ | $-\pi/2$ | $\pi/2\Gamma$ |

**SM-4**

The system shown in Fig. 5(b) can act as an entangled photon source when the QD is utilized as a single photon source. This requires the following three steps: (i) The QD is excited by a π-pulse from the outside and single photons are introduced into each cavity. (ii) The CNOT operation is performed, which entangles the photons trapped in the cavities. (iii) Photons are released into the waveguides of the external ports. Here, the operation in step (ii) has already been explained and step (iii) can be achieved by dynamic control of the $Q$-factors [S6, S7]. Therefore, we focus here on step (i).

First, we discuss the way in which single photons are introduced into the fifth and sixth cavities (Fig. 2). In this case, the $|x\rangle$ state in the QD is excited by the x-polarized π-pulse from the outside with all elements decoupled. Then, the QD is coupled with the fifth (sixth) cavity, where resonant Rabi oscillations are induced. Therefore, a single photon can successfully be trapped in the fifth (sixth) cavity after a time $\pi/4\Gamma$, for which the operation is the same as those in steps (A) and (C) in the q-CNOT gate.

It is difficult to apply this idea for the introduction of single photons into the third and fourth cavities because the resonant condition cannot be obtained for the Rabi oscillations due to the specific design of the q-CNOT gate. In other words, the coupling between the QD and each cavity is always achieved with inevitable detuning because the photon emission rates of the QD are $4\Gamma$ whereas those of the third and fourth cavities are designed to be $2\Gamma$.

In order to avoid this difficulty, a state of $(|1\rangle_{c3}|0\rangle_{c4} + |0\rangle_{c3}|1\rangle_{c4})|g\rangle/\sqrt{2} = |+\rangle_{L1}|g\rangle$ is first constructed by

coupling the QD and the third and fourth cavities simultaneously after the |y⟩ state is excited by the y-polarized π-pulse. This is achieved by the same phase differences as step (B) in Table SII, for which the evolution of the system can be described by $\hat{U}_2^t$:

$$\hat{U}_2^t |0\rangle_{c3}|0\rangle_{c4}|0\rangle_{c5}|0\rangle_{c6}|y\rangle = \frac{\exp(-i\omega_0 t)}{\sqrt{2}}\left\{\frac{1-\exp(-i4\Gamma t)}{\sqrt{2}}|+\rangle_{L1}|g\rangle \right.$$

$$\left. + \frac{1+\exp(-i4\Gamma t)}{\sqrt{2}}|0\rangle_{c3}|0\rangle_{c4}|y\rangle\right\}|0\rangle_{c5}|0\rangle_{c6}. \quad (S70)$$

Therefore, it can be confirmed that $\hat{U}_2^{\pi/4\Gamma}$ converts $|0\rangle_{c3}|0\rangle_{c4}|0\rangle_{c5}|0\rangle_{c6}|y\rangle$ into $|+\rangle_{L1}|0\rangle_{c5}|0\rangle_{c6}|g\rangle$. In this case, the state of the third and fourth cavities is positioned at the state of $|+\rangle_{L1}$ on the Bloch sphere, as shown in Fig. S3.

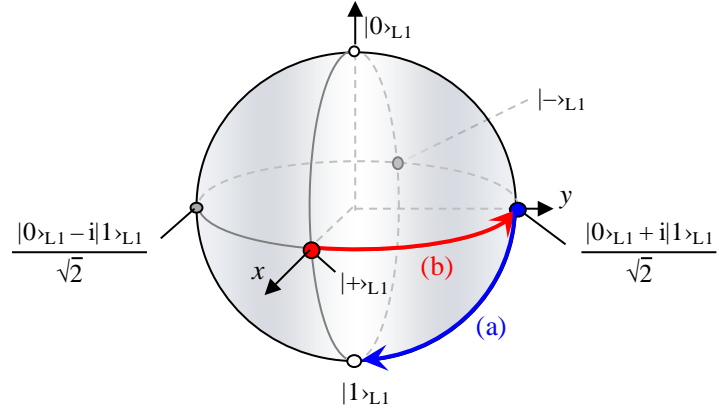

Figure S3. Trajectory on the Bloch sphere for introduction of a single photon into the third cavity.

Then, the state can be transformed into $|1\rangle_{c3}|0\rangle_{c4}|0\rangle_{c5}|0\rangle_{c6}|g\rangle$ ($= |1\rangle_{L1}|0\rangle_{c5}|0\rangle_{c6}|g\rangle$) or $|0\rangle_{c3}|1\rangle_{c4}|0\rangle_{c5}|0\rangle_{c6}|g\rangle$ ($= |0\rangle_{L1}|0\rangle_{c5}|0\rangle_{c6}|g\rangle$) by appropriate control of the phase differences, which are identical to the one-qubit operation: (a) the resonant Rabi oscillations and (b) control of the relative phase between $|1\rangle_{c3}|0\rangle_{c4}$ and $|0\rangle_{c3}|1\rangle_{c4}$. Here, the resonant condition can be obtained for the Rabi oscillation between $|1\rangle_{c3}|0\rangle_{c4}$ and $|0\rangle_{c3}|1\rangle_{c4}$ because the photon emission rates of the third and fourth cavities are the same ($2\Gamma$). The specific phase differences for operations (a) and (b) are summarized in Table SIII together with the operation that holds all elements decoupled. One can easily achieve actual control of the phase

differences, depending on the desired final state. For example, when a single photon should be introduced into the third cavity, $|+\rangle_{L1}|0\rangle_{c5}|0\rangle_{c6}|g\rangle$ is transformed into $(|0\rangle_{L1} + i|1\rangle_{L1})|0\rangle_{c5}|0\rangle_{c6}|g\rangle/\sqrt{2}$ by operation (b) with an interaction time of $\pi/2\Gamma$. Then, it evolves into $|1\rangle_{L1}|0\rangle_{c5}|0\rangle_{c6}|g\rangle$ by operation (a) after a time $\pi/4\Gamma$. As a result, a single photon can be introduced into the third cavity, as shown in Fig. S3. It is also possible to provide a single photon into the fourth cavity in a similar manner.

Thus, it has been shown that single photons can successfully be introduced into each cavity, on demand. It is straightforward to entangle these photons by the CNOT operation and to release them into the waveguides of the external ports, as described above.

Table SIII. Phase differences for the feeding of single photons into the third and fourth cavities.

| Operation | $\theta_{M3}$ | $\theta_{34}$ | $\theta_{4y}$ | $\theta_{My}$ | $\theta_{Mx}$ | $\theta_{5x}$ | $\theta_{56}$ | $\theta_{M6}$ |
|---|---|---|---|---|---|---|---|---|
| (a) | $\pi/2$ | $\pi$ | $-\pi/4$ | $\pi$ | $\pi$ | 0 | $\pi/2$ | $\pi$ |
| (b) | $-\pi/2$ | $5\pi/4$ | 0 | $\pi$ | $\pi$ | 0 | $\pi/2$ | $\pi$ |
| Hold | $\pi$ | $\pi/2$ | 0 | $\pi$ | $\pi$ | 0 | $\pi/2$ | $\pi$ |

## SM-5

Finally, we propose one possible structure for many qubits (Fig. S4); here, only one QD is used because it is difficult to prepare many QDs with identical transition energies using current technologies. The unit for the two-qubit gate in Fig. S4 is the same as that shown in Fig. 2 despite their seemingly different configurations. In this system, photons are initially provided from the single QD into the units for the one-qubit gates through waveguides, where dynamic control of the $Q$-factors [S6] and directional couplers are used. Each qubit is stored in the unit for the one-qubit gate. The one-qubit operations are then performed, whereas the two-qubit operations are realized only in the single unit for the two-qubit gate. Therefore, the photons of the control and target qubits should be transferred again before and after the two-qubit operations are carried out. After the computation, each photon is sent from the memory unit into the waveguides for readout and detected by single photon counters (not shown). In this structure, the initialization and the two-qubit gate operattions cannot be performed in parallel. However, it would be sufficient to demonstrate proof-of-principle experiments for several qubits or several tens of qubits. We believe that further development of the fabrication technology for many QDs (or alternative V-type three-level systems) with identical transition energies would allow our system to become more versatile in the future.

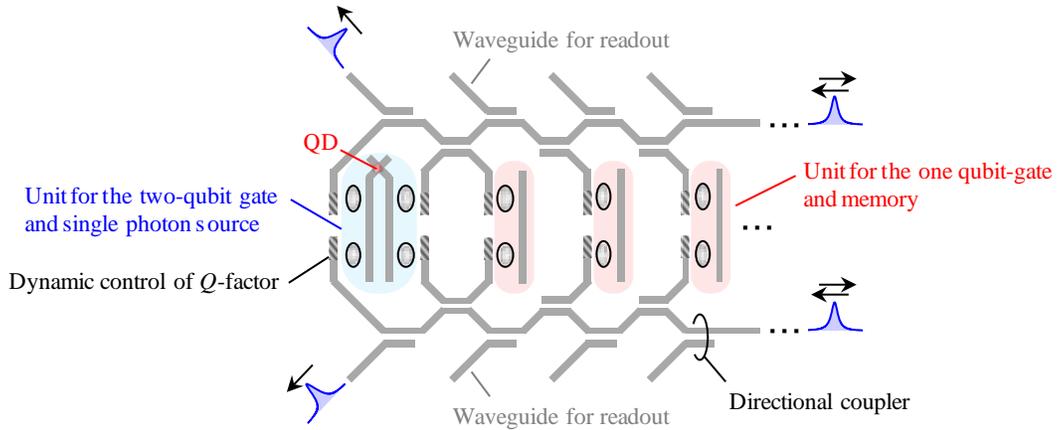

Figure S4. One possible structure for many qubits, using a single QD.

## Supplementary references